\documentclass[letterpaper, aps, tightenlines, nofootinbib, 11pt]{article}
\pdfoutput = 1

\usepackage{setspace}
\usepackage[bookmarks = false, colorlinks = true, linkcolor = blue, citecolor = purple]{hyperref}
\usepackage{shortcuts}
\usepackage{titlesec}
\usepackage{cite}
\usepackage{tocloft}

\usepackage{tikz}
\usetikzlibrary{arrows,decorations.pathmorphing,backgrounds,positioning,fit,petri}

\usepackage[margin = 2.5cm]{geometry}
\titleformat{\section}
  {\normalfont\large\bfseries}{\thesection}{1em}{}
\titleformat{\subsection}
  {\normalfont\large\it}{\thesubsection}{1em}{}
\titleformat{\subsubsection}
  {\normalfont\it}{\thesubsubsection}{1em}{}
\numberwithin{equation}{section}

\begin{document}

\begin{center}

\thispagestyle{empty}

\begin{flushright}
\texttt{SU-ITP-16/11}
\end{flushright}

\vspace*{5em}

{ \LARGE \bf Entanglement Entropy and Duality}

\vspace{1cm}

{\large \DJ or\dj e Radi\v cevi\'c}
\vspace{1em}

{\it Stanford Institute for Theoretical Physics and Department of Physics\\ Stanford University \\
Stanford, CA 94305-4060, USA}\\
\vspace{1em}
\texttt{djordje@stanford.edu}\\

\vspace{0.08\textheight}
\begin{abstract}
Using the algebraic approach to entanglement entropy, we study several dual pairs of lattice theories and show how the entropy is completely preserved across each duality. Our main result is that a maximal algebra of observables in a region typically dualizes to a non-maximal algebra in a dual region. In particular, we show how the usual notion of tracing out external degrees of freedom dualizes to a tracing out coupled to an additional summation over superselection sectors. We briefly comment on possible extensions of our results to more intricate dualities, including holographic ones.
\end{abstract}
\end{center}

\newpage

%{\small
%\tableofcontents
%}

\section{Introduction}

Entanglement entropy quantifies the amount of information about a quantum state that is lost upon restriction to a subsystem. Typically, by ``subsystem'' one means a spatial region, but in general it can be any subalgebra of observables that belong to the theory in question. This algebraic point of view has received increased interest in the context of defining entanglement entropy in gauge theories \cite{Casini:2013rba, Casini:2014aia, Radicevic:2014kqa, Radicevic:2015sza, Soni:2015yga, Ghosh:2015iwa, Aoki:2015bsa}, but the tools unearthed in this body of work can also be used to understand another deep question: the relation between entanglement entropy and field-theoretic dualities.

The crux of the algebraic approach to entanglement lies in constructing the density matrix $\rho_V$ associated to any subalgebra $\A_V$ of observables \cite{Ohya:2004, Balachandran:2013}. (We will often use ``subsystem'' to refer either to the rule $V$ that picks out the subalgebra, or to the subalgebra $\A_V$ itself.) This density operator is the unique element of $\A_V$ that is positive semi-definite, has unit trace in some natural representation, and reproduces the expectation values of all operators in the subsystem via $\avg \O = \Tr(\rho_V \O)$. The entanglement entropy of $\A_V$ is then defined as the von Neumann entropy $S = - \Tr(\rho_V \log \rho_V)$. When $V$ is a spatial region and $\A_V$ is the maximal algebra of observables in $V$, the density matrix defined in this algebraic way coincides with the one obtained by tracing out the degrees of freedom outside $V$ in the density matrix for the whole system. However, even if these conditions are not fulfilled, $\rho_V$ is still a legitimate density operator, and its entropy reflects the fact that certain measurements on the system are not accessible to us.

A general approach like this is indispensable when studying dualities. Typically, when two theories map to each other, there is a small mismatch between a theory that respects duality and a theory that we are used to dealing with, and the subalgebras dual to each other are not necessarily maximal algebras on any spatial region. For example, when a photon is dualized to a scalar in $d = 2$ spatial dimensions, the scalar zero mode obeys a certain compactness condition. At weak gauge theory coupling, the zero mode operator is absent altogether. The lack of this operator in the strongly coupled (``ordered'') regime of the scalar theory leads to both an area law and a term that corresponds to the topological entanglement entropy in the dual weakly coupled gauge theory \cite{Agon:2013iva}.

One upshot of this discussion is that there is no unique notion of entanglement entropy associated to a spatial region. Instead, for each region one can define a multitude of algebras associated to it, and each algebra choice comes with its own entanglement entropy. The maximal algebra is a natural choice, and indeed most of the current intuition comes from the entropy associated to this algebra, via the tracing out procedure. Nevertheless, this is still just a choice, and other algebras associated to a region --- for instance, one that differs from the maximal one by only one generator --- lead to (in principle) different measures of entanglement. We emphasize that this is not a mere UV ambiguity in the definition of the entropy (see also \cite{Casini:2013rba}), in the same way that a choice of boundary conditions in a path integral is not merely a UV effect. This analogy is not accidental; we will demonstrate below that a particular non-maximal choice of subalgebra can be represented as a tracing out of a full density matrix while summing over boundary conditions at the entangling edge.

The purpose of this paper is to study simple Ising systems on a lattice and to very explicitly show how the same entanglement entropy is exhibited on both sides of various dualities. In particular, we will focus on Kramers-Wannier (KW) dualities of the Ising model in $d = 1$ and $d = 2$ \cite{Savit:1979ny}, and on Jordan-Wigner/bosonization dualities of the Ising model in $d = 1$ dimensions. Working with the Ising model affords us a great degree of transparency, but our conclusions generalize to KW dualities of other Abelian theories in different dimensions.

\section{Entropy in a single spin}

Before sinking our teeth into pairs of dual theories, let us first warm up using a rather trivial example. We will use notation that immediately generalizes to more complicated cases. Consider a system consisting of a single spin with a two-dimensional Hilbert space $\H$. The algebra of Hermitian operators acting on this spin is $\A = \{\sigma^\mu\}$, where $\sigma^0 = \1$ and the remaining three operators, $\sigma^x$, $\sigma^y$, and $\sigma^z$, satisfy the commutation relations of the usual Pauli matrices. This algebra is generated by two operators, say $\sigma^x$ and $\sigma^z$; other operators are obtained as products of these two. There exist four subalgebras of $\A$:
\bel{
  \{\1\},\ \{\1,\,\sigma^x\},\ \{\1,\,\sigma^y\},\ \trm{and}\ \{\1,\,\sigma^z\}.
}
Other subsets of $\A$ are not algebras because they are not closed under multiplication.

Each of the above algebras has an associated reduced density matrix. For a subalgebra $\A_V = \{\1,\O\}$, the general density operator can be written as $\rho_V = \rho_\1 \1 + \rho_\O \O$, where $\rho_\1$ and $\rho_\O$ are numbers. These coefficients are uniquely determined by solving a system of two linear equations coming from the requirement
\bel{\label{def rho}
  \Tr(\rho_V \O) = \avg \O.
}
This uniqueness persists in more complicated examples with arbitrarily large algebras. For the trivial subalgebra $\A_V = \{\1\}$, the reduced density matrix is $\rho_V = \rho_\1 \1$, where $\rho_\1$ is chosen to ensure the operator has unit trace, in agreement with \eqref{def rho}.

The coefficients in the expansion
\bel{\label{expansion}
  \rho_V = \sum_{\O \in \A_V} \rho_\O \O
}
will all depend on the representation of the trace on the l.h.s.~of \eqref{def rho}. In principle, we can choose \emph{any} representation, and we would get a legal density matrix. The von Neumann entropy associated to $\rho_V$ does depend on this choice. For instance, the trivial algebra $\A_V = \{1\}$ can be represented as the identity operation on a space of arbitrary dimension $D$, and the entropy would then be $\log D$. There is no reason to believe that one representation is more fundamental than another; each is a different yardstick for measuring the entropy. In this paper we will always employ the natural choice that comes from the original Hilbert space on which the full algebra $\A$ was defined, and when comparing entropies of different algebras we will make sure to only compare the entropies associated to representations of the same dimensionality.

With this comment in mind, we choose to represent the operators $\1$ and $\O$ as $2\times 2$ matrices acting on vectors in $\H$. If $\A_V = \{\1\}$, the reduced density matrix is $\rho_V = \frac12\1$ and the entropy is $S_V = \log 2$ regardless of the original state of the system. This is natural, as having access only to the identity operator means that we have no way of measuring anything about the system, so we can do no better than to express it as a completely mixed state.

If $\A_V = \{\1, \sigma^z\}$, say, things are more interesting. If $\avg{\sigma^z} = 0$, the reduced density matrix is again $\rho_V = \frac12 \1$, describing a mixed state since we have no information whether the system is in the $+1$ or $-1$ eigenstate of one of the other two operators. The entropy is again $\log 2$. If $\avg{\sigma^z} = 1$, however, the reduced density matrix describes a pure state, $\rho_V = \qvec{\!\!\uparrow}\qvecconj{\uparrow\!\!}$, and the entropy is zero; this time the observable algebra is enough to determine all information about the state of the system.

The setup described so far has a very nice property that generalizes to all $\Z_2$ models we study in this paper: all operators except for the identity have zero trace, and all operators square to the identity. This allows us to multiply both sides of \eqref{expansion} with an operator $\O$ and take the trace, getting
\bel{\label{coefficients}
  \rho_\O = \frac{\avg\O}{\Tr\, \1}.
}
If we know the coefficients $\rho_\O$ of the full density matrix, we automatically know all the reduced density matrices: we just project the sum $\rho = \sum_{\O \in \A} \rho_\O \O$ to $\rho_V = \sum_{\O \in \A_V} \rho_\O \O$. In other examples, we may want to express the operators in $\A_V$ as acting on a smaller Hilbert space $\H_V$, in which case we need to restrict the sum to $\O \in \A_V$ \emph{and} to rescale all the surviving coefficients $\rho_\O$ by $\frac{\dim \H}{\dim \H_V}$. Doing this for a maximal algebra on a spatial subset $V$ gives the reduced matrix $\rho_V = \Tr_{\bar V} \rho$.

\section{Ising-Ising duality ($d = 1$)}

Consider the quantum Ising model defined on a chain with $L$ sites. Any conceivable operator in this model can be written as $\sigma^{\mu_1}_1 \ldots \sigma^{\mu_L}_L$, where $\sigma^\mu_i$ is a Pauli matrix acting on site $i$.\footnote{For clarity, we omit the $\otimes$ symbols and factors of $\1$. In proper notation, $\sigma_i^\mu$ would be written as $\1_1 \otimes \ldots \otimes \1_{i - 1} \otimes \sigma^\mu_i \otimes \1_{i + 1} \otimes \ldots \otimes \1_L$.} The version of the Ising model that possesses a Kramers-Wannier (KW) dual does not contain all of these operators; the needed algebra is generated by operators $\sigma^z_i$ and $\sigma^x_i \sigma^x_{i + 1}$ for $i = 1, \ldots, L - 1$. This choice reflects the adoption of open boundary conditions ($\sigma^z$ cannot be measured at the edges of the system).\footnote{There exists a version with a $\sigma^z$ at only one edge, but it would give us the same results as this one.} The Hamiltonian is
\bel{
  H = -\sum_{i = 1}^{L-1} \sigma^x_i \sigma^x_{i + 1} + h \sum_{i = 2}^{L - 1} \sigma^z_i,
}
and the Hilbert space $\H$ is taken to be $2^{L - 1}$-dimensional, with the spin on site 1 always being in state $\qvec +$ that satisfies $\sigma^x_1 \qvec + = \qvec +$.
At strong coupling ($h \gg 1$), there are two orthogonal ground states corresponding to two different boundary conditions,
\bel{
  \qvec{\Omega_\downarrow^{h \gg 1}} = \qvec{+\!\!\downarrow\downarrow\ldots\downarrow\downarrow}, \quad \qvec{\Omega_\uparrow^{h \gg 1}} = \qvec{+\!\!\downarrow\downarrow \ldots \downarrow\uparrow},
}
with $\sigma^z \qvec{\!\!\uparrow} = \qvec{\!\!\uparrow}$ and $\sigma^z \qvec{\!\!\downarrow} = -\qvec{\!\!\downarrow}$ as usual. At weak coupling ($h \ll 1$) the unique ground state is
\bel{
  \qvec{\Omega^{h \ll 1}} = \qvec{++\ldots+}.
}

The KW dual of this system is an Ising model defined on the links of the above chain. We define the dual algebra via generators $\tau^z_{i,\, i+1} = \sigma^x_i \sigma^x_{i + 1}$ and $\tau^x_{i - 1,\, i} \tau^x_{i,\, i + 1} = \sigma^z_i$, so the Hamiltonian of the dual space is
\bel{\label{1d dual H}
  H = - \~h \sum_{i = 1}^{L-1} \tau^z_{i,\, i+1} +  \sum_{i = 2}^{L - 1} \tau^x_{i-1,\,i} \tau^x_{i,\,i+1}, \quad \~h = \frac1h.
}
The $2^{L - 1}$-dimensional Hilbert space of Ising spins on links is isomorphic to the original Hilbert space, and we will denote its elements with $|\cdot\}$. The natural mappings between the two spaces map ground states to each other, and in terms of basis elements they are
\algnl{
  \begin{split}
    \qvec{\pm}_i \qvec{\pm}_{i + 1} \mapsto |\!\uparrow\}_{i,\,i+1},&\quad
    \qvec{\pm}_i \qvec{\mp}_{i + 1} \mapsto |\!\downarrow\}_{i,\,i+1}. \\
    |\pm\}_{i-1,\,i}|\pm\}_{i,\,i+1} \mapsto \qvec{\!\downarrow}_i,&\quad
    |\pm\}_{i-1,\,i}|\mp\}_{i,\,i+1} \mapsto \qvec{\!\uparrow}_i.
  \end{split}
}
This implements the standard picture of spin flips being dualized to kinks/domain walls by a KW transformation.

The lack of individual $\tau^x$ operators in the dual picture means that there is no measurement that would distinguish between states related by a global spin flip in the $\tau^x$ eigenbasis. This is not so in the original picture, where the first spin is fixed to be in the $\qvec +$ state, so all the other individual $\sigma^x_i$ eigenvalues can be measured, and there is no lack of information on the overall spin flip in the $\sigma^x$ basis. Rather, the global spin-flip symmetry of the dual model corresponds to the $\qvec{\!\uparrow}_L \mapsto \qvec{\!\downarrow}_L$ symmetry in the original model (note that $\sigma_L^z$ is also not an observable).

Let us now study entanglement entropy on both sides of the duality. Consider first a set of neighboring sites $V$ in the original picture. Assuming that $V$ is away from the edges of the system, the maximal algebra $\A_V$ supported on $V$ is generated by $|V|$ operators $\sigma^z_i$ and $|V| - 1$ operators $\sigma^x_i \sigma^x_{i + 1}$. The entanglement entropy that we wish to compute is the von Neumann entropy of the matrix $\rho_V = \sum_{\O \in \A_V} \rho_\O \O$ represented as an operator on the Hilbert space $\H_V$ of spins in $V$. The coefficients in this expansion are, according to \eqref{coefficients},
\bel{
  \rho_\O = \frac{\avg \O}{\dim \H_V}.
}

At strong coupling, the reduced density operator $\rho_V^{h \gg 1}$ is built out of all the operators with nonzero expectation values in the state $\qvec{\Omega_\downarrow^{h\gg 1}}$. (The result will be the same in the other ground state, of course.) The operators with nonzero vevs are all possible products of $\sigma^z$'s, and the density matrix is
\bel{\label{1d rho strong}
  \rho_V^{h \gg 1} = \frac1{2^{|V|}}\left(\1 - \sum_{i \in V} \sigma_i^z + \sum_{i < j} \sigma^z_i \sigma^z_j - \ldots + (-1)^{|V|} \prod_{i \in V} \sigma^z_i \right).
}
These matrices are all diagonal in the $\sigma^z$ eigenbasis, and it takes a simple counting exercise to determine that all the diagonal entries except for $\qvec{\!\downarrow\ldots\downarrow}\qvecconj{\downarrow\ldots\downarrow\!}$ are zero. Thus, $\rho_V^{h \gg 1}$ describes a pure state and the entanglement entropy at strong coupling is
\bel{
  S^{h \gg 1}_V = 0.
}

At weak coupling, the situation is inverted. The only operators with nonzero vevs are products of $\sigma^x_i \sigma^x_{i + 1}$, and all the vevs are equal to one.  The reduced density operator is
\bel{
  \rho^{h \ll 1}_V  = \frac1{2^{|V|}}\left(\1 + \sum_{i < j} \sigma^x_i \sigma^x_j + \sum_{i < j < k < l} \sigma^x_i \sigma^x_j \sigma^x_k \sigma^x_l + \ldots  \right).
}
Like before, it is sufficient to work in the $\sigma^x$ basis and count the $\pm1$ terms on the diagonal. The result is that matrix elements at positions $\qvec{+\ldots+}\qvecconj{+\ldots+}$ and $\qvec{-\ldots-}\qvecconj{-\ldots-}$ will each equal $1/2$, while all others will be zero. The weak-coupling reduced density matrix thus represents a mixed state with entropy
\bel{
  S^{h \ll 1}_V = \log 2.
}

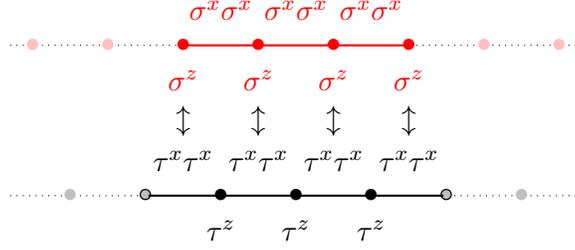
\begin{figure}[tb!]
\begin{center}

\begin{tikzpicture}[scale = 2]

  \draw[step = 0.5, dotted] (-1.9, 0) -- (1.9, 0);
  \draw[step = 0.5, dotted] (-1.9, -1) -- (1.9, -1);

  \foreach \x in {-1.75, -1.25, ..., 1.75}
    \draw (\x, 0) node[pink] {$\bullet$};

  \foreach \x in {-1.5, -1, ..., 1.5}
    \draw (\x, -1) node[lightgray] {$\bullet$};

  \draw[thick] (-0.98, -1) -- (0.98, -1);
  \draw (-1, -1) node {$\circ$};
  \draw (1, -1) node {$\circ$};

  \draw[thick, red] (-.73, 0) -- (.73, 0);

  \foreach \x in {-0.5, 0, 0.5} {
    \draw (\x, -1) node {$\bullet$};
    \draw (\x, -1.1) node[anchor = north] {$\tau^z$};
    \draw (\x, 0.1) node[anchor = south, red] {$\sigma^x \sigma^x$};
  };

  \foreach \x in {-0.75, -0.25, 0.25, 0.75} {
    \draw (\x, 0) node[red] {$\bullet$};
    \draw (\x, -0.9) node[anchor = south] {$\tau^x \tau^x$};
    \draw (\x, -0.1) node[anchor = north, red] {$\sigma^z$};
    \draw (\x, -0.5) node {$\updownarrow$};
  };

\end{tikzpicture}

\end{center}
\caption{\small \textsc{(color online)} KW duality in $d = 1$. Above: the original picture. Thick red dots are the set $V$, and operators generating its maximal algebra are explicitly labeled.  Below: dual picture. Black circles denote edge sites without $\tau^z$ operators; all operators in the dual algebra are also labeled.}
\label{fig 1d}
\end{figure}

In the dual picture, the algebra $\A_V$ maps to the non-maximal algebra $\~\A_{\~V}$ on the region $\~V$ with $|V| + 1$ sites (see Fig.~\ref{fig 1d}). This dual algebra is generated by operators $\tau^z_i$ for $i \in \~V - \del \~V$ and $\tau^x_i \tau^x_{i + 1}$ for $i \in \~V$; in other words, $\~\A_{\~V}$ is obtained by removing the edge operators $\tau^z_i$ from the maximal algebra on $\~V$. Note that the dimension of the Hilbert space on $\~V$ is different from the dimension of the corresponding Hilbert space on $V$. As discussed in the Introduction, this means that the entanglement entropy that is naturally calculated in the dual picture can be greater than the entropy in the original picture, as the algebras of the observables are the same but their representations differ. In order to meaningfully compare entropies on both sides of the duality, we will take dual operators to act on the $2^{|\~V|}$-dimensional Hilbert space $\H_{\~V}$ that has the first spin (at one edge of $\~V$) fixed to $|+\}$. This parallels the need to choose the spin on the end of the chain to be in the $\qvec+$ state.

Direct calculation can verify that the entanglement entropies are the same, as they should be since the dual subalgebras are isomorphic and the representations have the same dimension. It is instructive to see how this works out. The ground states at weak dual coupling are
\bel{
  |\Omega_1^{\~h \ll 1}\} = |+-+\ldots\},\quad |\Omega_2^{\~h \ll 1}\} = |-+-\ldots\},
}
and operators with nonzero vevs in these states are $\tau_i^x \tau_{i+1}^x$ and their products. The reduced density matrix $\rho_{\~V}^{\~h \ll 1}$ is found to be the pure state matrix $|+-+\ldots\}\{+-+\ldots|$ in both ground states, and the entanglement entropy is
\bel{
  S_{\~V}^{\~h \ll 1} = 0.
}
At strong dual coupling, the ground state is
\bel{
  |\Omega^{\~h \gg 1}\} = |\!\uparrow\ldots \uparrow\},
}
and the operators with nonzero vevs are $\tau^z_i$ and their products. With the inclusion of the edge operators, the reduced density matrix on $\H_{\~V}$ takes the form
\bel{\label{1d rho dual strong}
  \rho_{\~V}^{\~h \gg 1} = \frac1{2^{|\~V|}} \left(\1 + \sum_{i \in \~V - \del \~V} \tau^z_i + \ldots \right) \otimes \1.
}
The matrix in the parentheses is a pure state density matrix, $|\!\!\uparrow\ldots\uparrow\}\{\uparrow\ldots\uparrow\!\!|$. The entire von Neumann entropy of $\rho_{\~V}^{\~h \gg 1}$ comes from the identity operator at the edge where no boundary condition has been imposed:
\bel{
  S_{\~V}^{\~h \gg 1} = \log 2.
}

These calculations show that the naturally defined entanglement entropies of dual systems are
\bel{
  S_{\~V}^{\~h} = S_V^{h}.
}
While this may seem like a foregone conclusion given that the density matrices are evidently mapped to each other by duality, we point out that the origin of the entropy (when present) is different on the two sides of duality. In the original picture, the entropy came from mixing of the two states related by a global spin flip; in the dual picture, the same entropy came from the edge mode alone.  This is a very simple example of UV/IR correspondence engendered by duality.

\section{Ising-gauge duality ($d = 2$)}

\subsection{Setup}

Let us now define a $\Z_2$ gauge theory on a square $L\times L$ lattice. Operators and states are defined on links $\ell$, and the most general operator has the form $\prod_\ell \sigma^{\mu_\ell}_\ell$. Gauge-invariant states are those elements of the full Hilbert space $\H_0$ that are invariant under Gauss operators $G_i = \prod_\mu \sigma^x_{(i,\, \mu)}$ at any site $i$, with the product over all directions $\mu$ of links emanating from $i$. These states form the gauge-invariant Hilbert space $\H$. Operators that map $\H$ to $\H$ form the gauge-invariant algebra $\A$, which is generated by all the operators $\sigma^x_\ell$ and by products $W_p = \prod_{\ell \in p} \sigma^z_\ell$ around each plaquette $p$. The $W_p$ are magnetic operators (Wilson loops) and they create closed loops of electric flux. The $\sigma^x_\ell$ are electric operators and they create pairs of vortices (``magnetic flux insertions'').

As done for the Ising model, in order to define a theory with a KW dual, we choose that $\A$ has no generators on the edge of the lattice. In other words, we regard an electric operator $\sigma^x_\ell$ as unphysical/unobservable if $\ell$ does not belong to exactly two plaquettes. Now all the remaining gauge-invariant operators can be mapped to operators in the Ising model on the dual lattice via
\bel{
  W_p = \tau_p^z, \quad \sigma^x_\ell = \tau^x_p \tau^x_q,
}
with $p$ and $q$ being two plaquettes that both contain the link $\ell$. As before, we see that the system with no operators at its edges gets dualized to a system with no individual $\tau^x$ operator on any site.

\begin{figure}[tb!]
\begin{center}

\begin{tikzpicture}[scale = 2]

  \foreach \x in {-1, -0.5,..., 1} \draw (\x, -1) node[lightgray] {$\bullet$};
  \foreach \x in {-1, -0.5,..., 1} \draw (\x, 1) node[lightgray] {$\bullet$};
  \foreach \x in {-1, -0.5,..., 1} \draw (1, \x) node[lightgray] {$\bullet$};
  \foreach \x in {-1, -0.5,..., 1} \draw (-1, \x) node[lightgray] {$\bullet$};

  \foreach \x in {-0.75, -0.25, ..., 0.75} {
      \draw (-1.25, \x) node[red] {$\circ$};
      \draw (1.25, \x) node[red] {$\circ$};
      \draw (\x, -1.25) node[red] {$\circ$};
      \draw (\x, 1.25) node[red] {$\circ$};
      };

  \foreach \x in {-0.75, -0.25,..., 0.75}
    \foreach \y in {-0.75, -0.25,..., 0.75}
      \draw (\y, \x) node[red] {$\bullet$};

  \draw[step = 0.5, red, yshift = -0.25cm] (-1.23, -0.75) grid (1.23, 1);
  \draw[step = 0.5, red, xshift = -0.25cm] (-0.75, -1.23) grid (1, 1.23);
  \draw[step = 0.5, dotted] (-1.9, -1.9) grid (1.9, 1.9);
  \draw[step = 0.5, thick] (-1, -1) grid (1, 1);

\end{tikzpicture}

\end{center}
\caption{\small \textsc{(color online)} Duality in $d = 2$: Thick black lines are the set $V$, and grey circles denote edge sites in $\del V$. Electric operators $\sigma^x$ are defined on all thick black lines, and magnetic operators $W$ are defined on all thick black plaquettes. All red circles together form the dual set $\~V$. Operators $\tau^z$ on filled-in circles belong to the dual algebra $\~\A_{\~V}$, as do all $\tau^x\tau^x$ pairs on sites connected by red lines. }
\label{fig 2d}
\end{figure}
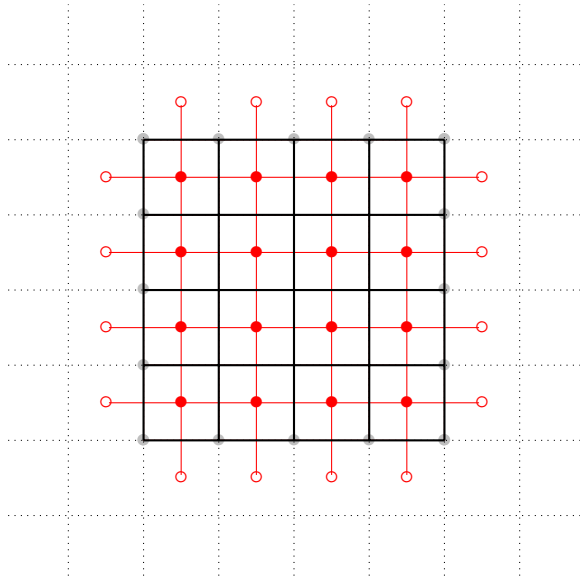

The mapping of Hilbert spaces presents more subtleties than in $d = 1$. The Hilbert space $\~\H$ of the Ising model has $2^{(L-1)^2}$ dimensions, as it is a product of two-dimensional Hilbert spaces on each plaquette. The gauge-invariant Hilbert space $\H$ has the same dimension, but the full space $\H_0$ has a much greater dimension, $2^{2L(L - 1)}$, being a product of two-dimensional spaces on each link. The full Hilbert space cannot be mapped to the Ising model; only gauge-invariant states map. Since the entanglement entropy in gauge theories is typically (if tacitly) calculated using the full Hilbert space \cite{Gromov:2014kia, Kitaev:2005dm, Hamma:2005zz, Levin:2006zz, Radicevic:2014kqa, Radicevic:2015sza, Soni:2015yga, Pretko:2015zva, Aoki:2015bsa, Buividovich:2008gq, Buividovich:2008yv, Casini:2013rba, Casini:2014aia, Donnelly:2011hn, Donnelly:2012st, Donnelly:2014gva, Ghosh:2015iwa}, we might expect large differences between entanglement entropies that are naturally calculated on the two sides of KW duality. This does not happen: the inclusion of Gauss law operators $G_i$ in the algebra of observables effectively projects the full Hilbert space down to the physical one.  We will later explain how this works in detail.

We take the gauge theory Hamiltonian to be
\bel{
  H = g \sum_\ell \sigma^x_\ell - \sum_p W_p.
}
The strong coupling ground state is degenerate, just like it was for the Ising chain. There are two physical ground states. Both contain a product of states $\qvec -$ at each interior link, and they differ in how the Gauss law is realized at the edge of the system. (Each of these states is obtained from the other by adding an (unobservable) electric flux loop along the system edge.) These states confine electric fields, and excitations are loops of electric flux. At weak coupling, the ground state $\qvec{\Omega^{g \ll 1}}$ has $W_p = 1$ on each plaquette. This is satisfied by $\prod_\ell \qvec{\!\!\downarrow}_\ell$ and by all other states obtained by acting on this one with products of $G_i$. The only gauge-invariant ground state is the sum of all of these states, and this is $\qvec{\Omega^{g \ll 1}}$. This ground state can also be expressed as the unweighted sum over all possible electric flux loop excitations of either strong coupling state $\qvec{\Omega^{g \gg 1}_{1/2}}$.

The dual Hamiltonian is
\bel{
  H = \sum_{\avg{p,\, q}} \tau^x_p \tau^x_q - \~g\sum_p \tau^z_p, \quad \~g  = \frac1g.
}
which is just the higher-dimensional analogue of \eqref{1d dual H}. As before, we denote the dual quantum states by $|\cdot\}$. At strong dual coupling, the ground state is
\bel{
  |\Omega^{\~g\gg 1}\} = \prod_p |\!\uparrow\}_p,
}
and at weak dual coupling the two ground states are the two possible ``checkerboard'' tilings of the lattice by $|+\}$ and $|-\}$ states,
\bel{
  |\Omega_1^{\~g \ll 1}\} = \prod_{i,\, j = 1}^{L - 1} |(-1)^{i + j}\}_{(i,\, j)},\quad |\Omega_2^{\~g \ll 1}\} = \prod_{i,\, j = 1}^{L - 1} |(-1)^{i + j + 1}\}_{(i,\, j)}.
}

\subsection{Entanglement in gauge theory}

The entanglement entropy on the gauge theory side has been calculated for both strong and weak coupling \cite{Soni:2015yga, Ghosh:2015iwa, Donnelly:2011hn, Radicevic:2015sza, Casini:2013rba}. In order to emphasize and further illustrate our operator approach to density matrices, we present a telegraphic derivation these well-known results. We will focus on the entropy associated to the maximal algebra that can be placed upon a set of links $V$.

At weak coupling, all products of Wilson loops satisfy $\avg{\prod_p W_p} = 1$, and all electric operators and their products have vanishing vevs --- with the exception of Gauss operators, $G_i$, whose products all satisfy $\avg{\prod_i G_i} = 1$. All these operators with nonzero vevs mutually commute. Thus, the reduced density operator for the algebra $\A_V$ can be written as
\bel{
  \rho_V^{g \ll 1} = \frac{1}{2^{\#\trm{links}(V)}} \left(\1 + \sum_p W_p + \sum_i G_i + \sum_{p \neq q} W_p W_q + \sum_{i \neq j} G_i G_j + \sum_{i,\, p} G_i W_p + \ldots \right),
}
where the denominator is the dimension of the Hilbert space of spins on all links in $V$ (without any regard for gauge invariance), and the sums include all Gauss operators and elementary Wilson loops that generate $\A_V$. This density operator can also be written as
\bel{
  \rho_V^{g \ll 1} =  \frac{2^{\#\trm{stars}(V)}}{2^{\#\trm{links}(V)}} \left(\1 + \sum_p W_p + \sum_{p \neq q} W_p W_q + \ldots\right) \prod_i \frac{\1 + G_i}2.
}
(Note that a Gauss operator $G_i$ is in $\A_V$ if and only if all links emanating from site $i$ are in $V$; such configurations of links are called ``stars.'') Each operator $\frac12(\1 + G_i)$ projects onto the space of states that obey the Gauss law at site $i$. This extremely convenient fact allows us to forget about the original set of degrees of freedom inside $V$ and to work just with gauge-invariant basis vectors. However, we must still work with gauge-variant degrees of freedom at edge sites of $V$, as the associated Gauss operators will not be in $\A_V$. This shows how the general operator prescription reduces to the one studied by extended Hilbert space and/or superselection sector techniques \cite{Soni:2015yga, Ghosh:2015iwa}.\footnote{This conclusion holds generally, not just in the ground state at weak coupling. Gauss operators belong to the center of $\A_V$, they always have unit expectation values for gauge-invariant states, and hence any density matrix can be written as a projector to the gauge-invariant subspace.}

The von Neumann entropy of $\rho_V^{g \ll 1}$ is easy to compute when the matrix is stripped of the projection operators and expressed in the basis that diagonalizes the Wilson loops. The density matrix is then diagonal and uniformly mixes the $2^{|\del V| - 1}$ different basis vectors that correspond to states with $W_p = 1$ and $G_i = 1$ at all plaquettes and stars. (Here we assume that $V$ does not contain disconnected components.) Its entropy takes the familiar form
\bel{
  S^{g \ll 1}_V = (|\del V| - 1) \log 2.
}

At strong coupling, the situation is somewhat simpler: operators with nonzero vevs are all products of $\sigma^x$'s with $\avg{\sigma^x_{\ell_1} \ldots \sigma^x_{\ell_n}} = (-1)^n$. Gauss operators are a special case of these, and do not need to be treated separately. The reduced density matrix is
\bel{
  \rho^{g \gg 1}_V = \frac1{2^{\#\trm{links}(V)}} \left(\1 - \sum_\ell \sigma^x_\ell + \sum_{\ell \neq \ell'} \sigma^x_\ell \sigma^x_{\ell'} - \ldots\right).
}
In the electric basis this matrix is diagonal and by simple inspection we see that the entry corresponding to the basis vector $\prod_{\ell \in V} \qvec{-}_\ell$ is equal to unity; therefore other entries must be zero, and the matrix is pure. This again reproduces the well-known result
\bel{
  S^{g \gg 1}_V = 0.
}

\subsection{Entanglement in the dual picture}

Just like in the previous Section, the maximal algebra on a set of links does not map to a maximal algebra on a set of sites on the dual lattice. Instead, the dual algebra $\~\A_{\~V}$ lacks individual $\tau^x$ generators on any site but contains an extra set of $\tau^x_p \tau^x_q$ generators that wrap the edge of the original region $V$ (see Fig.~\ref{fig 2d}).

At strong dual coupling, the operators with nonzero vev are $\tau_p^z$ and their products. Note that duals to Gauss operators, i.e.~products of $\tau^x_p \tau^x_q$ along closed contours, have vevs equal to $1$ not by virtue of the state being special, but rather purely algebraically, because each $\tau^x_p$ in that product is repeated an even number of times and the full product just gives the identity. These are not independent observables the way $G_i$'s were in the gauge theory.

The reduced density operator is thus
\bel{
  \rho_{\~V}^{\~g \gg 1} = \frac 1{2^{|\~V|}} \left(\1 + \sum_p \tau^z_p + \sum_{p \neq q} \tau^z_p \tau^z_q + \ldots + \prod_p \tau^z_p \right) \otimes \sideset{}{'}\prod_{p \in \del \~V} \1_p,
}
where the indices $p$ and $q$ in the parentheses run over the interior of the dual region, $\~V - \del\~V$. The product over edge degrees of freedom is primed to denote that it runs over $|\del \~V| - 1$ sites; this is because we need the representation of the dual subalgebra to have the same dimension as in the original picture, and hence we fix one boundary site to always be in the $|+\}$ state. This is a straightforward generalization of the $d = 1$ case \eqref{1d rho dual strong}, and the entropy is simply
\bel{
  S_{\~V}^{\~g \gg 1} = (|\del \~V| - 1) \log 2.
}
Since $|\del V| = |\del\~ V|$, the dual entropies $S_{\~V}^{\~g \gg 1}$ and $S_{V}^{g \ll 1}$ are equal, as they should be.

At weak dual coupling, the operators with nonzero expectations are $\tau^x_p \tau^x_q$ pairs and their products, \emph{except} for the products along closed loops of links. The reduced density matrix is
\bel{
  \rho_{\~V}^{\~g \ll 1} = \frac1{2^{|\~V|}} \left(\1 + \sum_{p\neq q} (-1)^{|p - q|} \tau^x_p \tau^x_q + \ldots \right).
}
By the same trick used in its dual strongly coupled gauge theory case, we notice that, in the $\tau^x$ eigenbasis, the vector with alternating $+$ and $-$ states is an element of $\H_{\~V}$ that gives a unit entry on the diagonal of $\rho_{\~V}^{\~g \ll 1}$. Since this is a density matrix, all other entries must be zero, and the entanglement entropy is
\bel{
  S_{\~V}^{\~g \ll 1} = 0.
}

We see that $S_{\~V}^{\~g} = S_V^{g}$ holds in $d = 2$, just like it did in $d = 1$. This time the interesting effect is the topological piece of the weak-coupling entropy, $-\log 2$, and the corresponding term in the strongly coupled scalar entropy. In the gauge theory this term is well-understood: due to the Gauss law, the total electric flux passing through the edge $\del V$ must be zero, this leads to a constraint on the types of states that the interior can be in, and therefore the entropy is smaller than the area law term that one may na\"ively expect. In the dual picture, the reduced density matrix uniformly mixes all edge modes (modulo an overall spin flip) even though the interior is in an ordered state. When computing an expectation value, this mixing can be implemented as a sum over all possible domain walls on a spin chain located on the entangling edge $\del \~V$. In the case at hand, there are $2^{|\del \~V| - 1}$ different domain walls.

This phenomenon is already known for the dual of $U(1)$ gauge theory \cite{Agon:2013iva}. The entanglement entropy of a compact scalar at small radius was found to come from the sum over configurations with different windings along the entanglement edge in the replica path integral. In the language of the present paper, the sum over winding sectors follows from the fact that the edge $\del\~V$ does not admit any position operators as observables. Adding these edge operators would project us to a sector with zero winding; equivalently stated, the reduced density matrix of this enlarged algebra would have to be that of a pure state in order to reproduce expectation values of the newly added operators. There exist related discussions in the contexts of $d = 3$ gauge theories \cite{Pretko:2015zva}, self-dual higher form gauge theories \cite{Ma:2015xes}, and the $d = 1$ Ising model \cite{Ohmori:2014eia}.

\section{Bosonization ($d = 1$)}

As our final example, we study the Jordan-Wigner transformation between the Ising chain and a system of Majorana fermions. This is a rather simple setup, but it will provide us with an example of a system where a nonlocal set of generators is needed to form the subalgebra of one side of the duality.

This time, we work with an Ising chain on $L$ sites with all operators present, and with Hamiltonian
\bel{
  H = -\sum_{i = 1}^{L-1} \sigma^x_i \sigma^x_{i + 1} + h \sum_{i = 1}^{L} \sigma^z_i.
}
The dual Majorana operators are
\bel{
  c_i = \sigma^x_i \prod_{j < i} \sigma^z_j ,\quad d_i = \sigma^y_i \prod_{j < i} \sigma^z_j .
}
These operators are Hermitian and any two nonidentical ones anticommute. Often, $d_i$ is written as $c_{i + 1/2}$. The dual Hamiltonian is
\bel{
  H = i \sum_{i = 1}^{L - 1} d_i c_{i + 1} - i h \sum_{i = 1}^L c_i d_i.
}
(Note that due to the anticommutation between all $c$'s and $d$'s, their products must be multiplied by $i$ to give Hermitian operators.) The Majoranas $c_i$ and $d_i$ act on the $2^i$-dimensional Hilbert space of a complex fermion at sites $1$ through $i$, which is in turn isomorphic to the Hilbert space of $i$ spins.

Let us consider the entropy associated to the algebra generated by a set of adjacent $c$'s and $d$'s on sites $V$. At strong coupling ($h \gg 1$), the ground state is a ``Majorana superconductor,'' a condensate of pairs of fermions that are coupled by the $h$ term in $H$. Individual Majorana fermions have vanishing vevs, and the only operators with nonvanishing expectations are the identity and products of pairs $i c_i d_i$. In the spin language, these are the $\sigma^z_i$ operators that detect that the ground state is ordered.

When choosing the representation of the reduced algebra $\A_V$, it is important to keep it the same dimension as in the original spin picture. Thus, even though all operators have trailing $\sigma^z$'s or $\1$'s going all the way to the beginning of the chain, we choose to represent all operators as matrices acting on the $2^{|V|}$-dimensional space of spins/fermions living in $V$. If the algebra $\A_V$ contains both Majoranas on each site in $V$, the strong coupling entropy is zero, in complete analogy with \eqref{1d rho strong}. If not, i.e.~if the system's right edge $V$ cuts between sites $i$ and $i + \frac12$, this is equivalent to removing $\sigma_i^z$ but leaving $\sigma^x_i$ in the subalgebra. If the system's left edge cuts between $i - \frac12$ and $i$, the situation is a bit more complicated because then both $\sigma^y_{i - 1}$ and $\sigma^x_{i - 1}$ will appear in the algebra, but the latter operator will only appear in products with other operators on sites $i$ and onwards.

At weak coupling, the spin system has two ground states (this time we have not imposed boundary conditions to lift this degeneracy). They are related by a global spin flip. In the Majorana picture, however, the degeneracy comes from the edge Majoranas $c_1$ and $d_L$ which are free (while all the others are paired up by the $d_i c_{i + 1}$ terms in $H$). If we pick the algebra $\A_V$ such that it doesn't split any of these pairs of Majoranas, we will get zero entropy. If, however, we pick the algebra such that it is a maximal algebra on a set of spin sites, then it will necessarily cut through two pairs of coupled Majoranas. The final result in this case is an entropy of $\log 2$. In the spin language this entropy came from the mixing of two states on $V$ related by a global spin flip, and in the fermion language it came from the dangling Majoranas at the edge of $V$. Once again, this is an example of a UV/IR connection.

\section{Outlook}

The main purpose of this paper was to provide a point of view from which entanglement entropy becomes an object that is naturally preserved by various dualities. The price we paid was the need to generalize entanglement entropy away from the usual ``tracing out'' procedure. From the point of view of this usual procedure, the generalized notion of entanglement entropy amounts to introducing summations over sectors labeled by eigenvalues of operators removed from a maximal algebra on a spatial region. In gauge theories, this summation over superselection sectors has been the subject of a lot of attention \cite{Casini:2013rba, Casini:2014aia, Radicevic:2014kqa, Radicevic:2015sza, Soni:2015yga, Ghosh:2015iwa, Aoki:2015bsa}, and in this paper we have shown that it can be understood to follow from excluding Gauss operators at the edges of the system from the observable subalgebra. In scalar theories, we have seen how excluding edge operators leads to a sum over winding sectors around the edge (or solitonic configurations on the edge), a variant of which was already considered in \cite{Agon:2013iva} in order to study entanglement in a scalar dual to a $d = 2$ Maxwell theory.

The entropy of a non-maximal algebra on a spatial region $V$ may seem undeserving of the name ``entanglement'' entropy. In particular, this entropy can be defined even in $d = 0$, as we did for the case of a single spin, and here there are no spatial regions to entangle. However, it seems that distinguishing between purely spatial entanglement entropy and these other entropies is not productive, as duality mixes up these notions. Moreover, at least in the case of Majorana fermions, there is even a way to map each generator of the Ising chain algebra to a Majorana operator on a separate spatial site, giving a direct geometric interpretation to each spin operator.

Requiring additional symmetries is a useful way to tame the multitude of subalgebras that can be placed on a spatial region. For instance, if we work on a lattice with spherical symmetry and we pick $V$ to be a ball, then requiring that $\A_V$ be spherically symmetric significantly restricts the set of allowed subalgebras, as we can now only remove a generator from all points at the edge. In fact, demanding enough symmetry may pick out a unique algebra (up to differences leading to nonuniversal terms only), as evidenced by the fact that supersymmetric Renyi entropies of certain superconformal theories on spherical regions (computed via localization of the replica trick path integral) agree across dualities without any manual summation over superselection sectors \cite{Nishioka:2013haa}.

Our results easily generalize to other Abelian theories. In particular, $\Z_k$ and $U(1)$ theories with known duals all follow the pattern of mapping a maximal algebra to a non-maximal one. In $d = 3$, where a gauge theory maps to another gauge theory via electric-magnetic duality, our results rather reassuringly imply that the maximal algebra on one side (the ``electric center'' choice) will map to the ``magnetic center'' choice on the other side of the duality. We have not touched upon dualities of nonabelian theories, as these are much more complex, but we expect that a similar story will hold.

A duality that we have so far not mentioned at all is holography. The algebraic approach to entanglement entropy in gravitational theories is still nascent \cite{Donnelly:2015hta, Donnelly:2016auv, Giddings:2015lla}, and even bulk operator reconstruction based on boundary data is very much an active field of research (see, for instance, \cite{Morrison:2014jha, Jafferis:2014lza, Hamilton:2006az, Kabat:2011rz, Dong:2016eik, Almheiri:2014lwa, Headrick:2014cta, Wall:2012uf, Czech:2012bh}). Nevertheless, holographic dualities qualitatively behave like the dualities studied in this paper: they are strong-weak coupling dualities, there exists a UV/IR connection \cite{Susskind:1998dq}, and entanglement entropy in the bulk appears to be equal to the one on the boundary, inasmuch as we know how to define entanglement in quantum gravity (see e.g.~\cite{Jafferis:2015del, Faulkner:2013ana}).  Based on this and on the intuition developed in this paper, a reasonable speculation at this stage would be that a maximal algebra in a subregion does not holographically map to a maximal algebra in a dual subregion, so any statement about the duality of subregions must be supplemented with rules about how to exclude certain operators (or how to sum over corresponding sectors in the path integral) on at least one side of the duality. It would be interesting to understand what boundary operators are missed by the construction of \cite{Dong:2016eik}, which proves that local bulk operators in the entanglement wedge of a boundary region $V$ are dual to boundary operators supported only in $V$. For example, local boundary operators in $V$ are dual to classical gravity backgrounds, i.e.~to solitons that live in the entire bulk, so we should expect that the algebra dual to local bulk operators in the entanglement wedge can contain local operators in the boundary only in some approximate sense. Making these speculations precise is a difficult but extremely rewarding task left for future work.

\section*{Acknowledgments}

It is a pleasure to thank Chao-Ming Jian, Jen Lin, Steve Shenker, and Itamar Yaakov for discussions. The author is supported by a Stanford Graduate Fellowship.

\end{document}